\pdfoutput=1
\documentclass[12pt,a4paper]{article}

\usepackage{ifthen} 
\newboolean{pdflatex}
\setboolean{pdflatex}{true} 

\newboolean{articletitles}
\setboolean{articletitles}{true} 

\newboolean{uprightparticles}
\setboolean{uprightparticles}{false} 

\def\paperauthors{LHCb collaboration} 
\def\paperasciititle{Tracking of charged particles with nanosecond lifetimes at LHCb} 
\def\papertitle{Tracking of charged particles with nanosecond lifetimes at LHCb} 
\def\paperkeywords{{High Energy Physics}, {LHCb}} 
\def\papercopyright{\the\year\ CERN for the benefit of the LHCb collaboration} 
\def\paperlicence{CC BY 4.0 licence}
\def\paperlicenceurl{https://creativecommons.org/licenses/by/4.0/}

\usepackage[top=1in, bottom=1.25in, left=1in, right=1in]{geometry}

\usepackage{microtype}
\usepackage{lineno}  
\usepackage{xspace} 
\usepackage{caption} 

\usepackage{graphicx}  
\usepackage{color}
\usepackage{colortbl}
\graphicspath{{./figs/}} 

\usepackage{amsmath} 
\usepackage{amssymb}
\usepackage{amsfonts}
\usepackage{upgreek} 

\newcommand*\patchAmsMathEnvironmentForLineno[1]{%
\expandafter\let\csname old#1\expandafter\endcsname\csname #1\endcsname
\expandafter\let\csname oldend#1\expandafter\endcsname\csname
end#1\endcsname
 \renewenvironment{#1}%
   {\linenomath\csname old#1\endcsname}%
   {\csname oldend#1\endcsname\endlinenomath}%
}
\newcommand*\patchBothAmsMathEnvironmentsForLineno[1]{%
  \patchAmsMathEnvironmentForLineno{#1}%
  \patchAmsMathEnvironmentForLineno{#1*}%
}
\AtBeginDocument{%
\patchBothAmsMathEnvironmentsForLineno{equation}%
\patchBothAmsMathEnvironmentsForLineno{align}%
\patchBothAmsMathEnvironmentsForLineno{flalign}%
\patchBothAmsMathEnvironmentsForLineno{alignat}%
\patchBothAmsMathEnvironmentsForLineno{gather}%
\patchBothAmsMathEnvironmentsForLineno{multline}%
\patchBothAmsMathEnvironmentsForLineno{eqnarray}%
}

\usepackage{hyperxmp}

\usepackage[pdftex,
            pdfauthor={\paperauthors},
            pdftitle={\paperasciititle},
            pdfkeywords={\paperkeywords},
            pdfcopyright={Copyright (C) \papercopyright},
            pdflicenseurl={\paperlicenceurl}]{hyperref}

\usepackage[colorinlistoftodos,textsize=scriptsize]{todonotes}

\usepackage[bottom,flushmargin,hang,multiple]{footmisc}

\usepackage[all]{hypcap} 

\usepackage{xspace} 
\usepackage{upgreek}

\def\lhcb   {\mbox{LHCb}\xspace}

\def\velo   {VELO\xspace}

\def\MagUp {\mbox{\em Mag\kern -0.05em Up}\xspace}

\ifthenelse{\boolean{uprightparticles}}%
{

 \def\Pmu         {\ensuremath{\upmu}\xspace}                 
 \def\Pnu         {\ensuremath{\upnu}\xspace}                 
                  
 \def\Ppi         {\ensuremath{\uppi}\xspace}

 \def\Ppsi        {\ensuremath{\uppsi}\xspace}

 \def\PDelta      {\ensuremath{\Delta}\xspace}                 
 \def\PXi         {\ensuremath{\Xi}\xspace}                 
 \def\PLambda     {\ensuremath{\Lambda}\xspace}                 
 \def\PSigma      {\ensuremath{\Sigma}\xspace}                 
 \def\POmega      {\ensuremath{\Omega}\xspace}                 
 \def\PUpsilon    {\ensuremath{\Upsilon}\xspace}
 \let\oldPi\Pi
 \def\PPi         {\ensuremath{\oldPi}\xspace}

 \def\PB      {\ensuremath{\mathrm{B}}\xspace}                 
                  
 \def\PD      {\ensuremath{\mathrm{D}}\xspace}

 \def\PJ      {\ensuremath{\mathrm{J}}\xspace}                 
 \def\PK      {\ensuremath{\mathrm{K}}\xspace}

 \def\Pb      {\ensuremath{\mathrm{b}}\xspace}                 
 \def\Pc      {\ensuremath{\mathrm{c}}\xspace}

 \def\Pi      {\ensuremath{\mathrm{i}}\xspace}

 \def\Pn      {\ensuremath{\mathrm{n}}\xspace}                 
                  
 \def\Pp      {\ensuremath{\mathrm{p}}\xspace}

 \def\Ps      {\ensuremath{\mathrm{s}}\xspace}

 \def\thebaroffset{0.0em}
}
{

 \def\Pmu         {\ensuremath{\mu}\xspace}                 
 \def\Pnu         {\ensuremath{\nu}\xspace}                 
                  
 \def\Ppi         {\ensuremath{\pi}\xspace}

 \def\Ppsi        {\ensuremath{\psi}\xspace}                 
                  
 \mathchardef\PDelta="7101
 \mathchardef\PXi="7104
 \mathchardef\PLambda="7103
 \mathchardef\PSigma="7106
 \mathchardef\POmega="710A
 \mathchardef\PUpsilon="7107
 \mathchardef\PPi="7105
                  
 \def\PB      {\ensuremath{B}\xspace}                 
                  
 \def\PD      {\ensuremath{D}\xspace}

 \def\PJ      {\ensuremath{J}\xspace}                 
 \def\PK      {\ensuremath{K}\xspace}

 \def\Pb      {\ensuremath{b}\xspace}                 
 \def\Pc      {\ensuremath{c}\xspace}

 \def\Pi      {\ensuremath{i}\xspace}

 \def\Pn      {\ensuremath{n}\xspace}                 
                  
 \def\Pp      {\ensuremath{p}\xspace}

 \def\Ps      {\ensuremath{s}\xspace}

 \def\thebaroffset{0.18em}
}
\newcommand{\offsetoverline}[2][\thebaroffset]{\kern #1\overline{\kern -#1 #2}}%

\makeatletter
\ifcase \@ptsize \relax
  \newcommand{\miniscule}{\@setfontsize\miniscule{4}{5}}
\or
  \newcommand{\miniscule}{\@setfontsize\miniscule{5}{6}}
\or
  \newcommand{\miniscule}{\@setfontsize\miniscule{5}{6}}
\fi
\makeatother

\DeclareRobustCommand{\optbar}[1]{\shortstack{{\miniscule (\rule[.5ex]{1.25em}{.18mm})}
  \\ [-.7ex] $#1$}}

\def\mup        {{\ensuremath{\Pmu^+}}\xspace}
\def\mun        {{\ensuremath{\Pmu^-}}\xspace} 

\def\neub       {{\ensuremath{\overline{\Pnu}}}\xspace}

\def\neumb      {{\ensuremath{\neub_\mu}}\xspace}

\def\squark    {{\ensuremath{\Ps}}\xspace}

\def\cquark    {{\ensuremath{\Pc}}\xspace}

\def\bquark    {{\ensuremath{\Pb}}\xspace}

\def\pion   {{\ensuremath{\Ppi}}\xspace}
\def\piz    {{\ensuremath{\pion^0}}\xspace}
\def\pip    {{\ensuremath{\pion^+}}\xspace}
\def\pim    {{\ensuremath{\pion^-}}\xspace}

\def\kaon    {{\ensuremath{\PK}}\xspace}

\def\KorKbar {\kern \thebaroffset\optbar{\kern -\thebaroffset \PK}{}\xspace}

\def\Kp      {{\ensuremath{\kaon^+}}\xspace}
\def\Km      {{\ensuremath{\kaon^-}}\xspace}

\def\KS      {{\ensuremath{\kaon^0_{\mathrm{S}}}}\xspace}

\def\D       {{\ensuremath{\PD}}\xspace}

\def\DorDbar {\kern \thebaroffset\optbar{\kern -\thebaroffset \PD}\xspace}

\def\Dp      {{\ensuremath{\D^+}}\xspace}
\def\Dm      {{\ensuremath{\D^-}}\xspace}

\def\DpDm    {\ensuremath{\Dp {\kern -0.16em \Dm}}\xspace}

\def\B       {{\ensuremath{\PB}}\xspace}

\def\BorBbar {\kern \thebaroffset\optbar{\kern -\thebaroffset \PB}\xspace}

\def\Bd      {{\ensuremath{\B^0}}\xspace}

\def\BdorBdbar {\kern \thebaroffset\optbar{\kern -\thebaroffset \Bd}\xspace}

\def\Bs      {{\ensuremath{\B^0_\squark}}\xspace}

\def\BsorBsbar {\kern \thebaroffset\optbar{\kern -\thebaroffset \Bs}\xspace}

\def\jpsi     {{\ensuremath{{\PJ\mskip -3mu/\mskip -2mu\Ppsi}}}\xspace}

\def\Y#1S{\ensuremath{\PUpsilon{(#1S)}}\xspace}

\def\proton      {{\ensuremath{\Pp}}\xspace}

\def\neutron     {{\ensuremath{\Pn}}\xspace}

\def\Lz          {{\ensuremath{\PLambda}}\xspace}

\def\LorLbar     {\kern \thebaroffset\optbar{\kern -\thebaroffset \PLambda}\xspace}

\def\Sigmares    {{\ensuremath{\PSigma}}\xspace}

\def\Sigmap      {{\ensuremath{\Sigmares{}^+}}\xspace}
\def\Sigmam      {{\ensuremath{\Sigmares{}^-}}\xspace}

\def\Xires       {{\ensuremath{\PXi}}\xspace}

\def\Xim         {{\ensuremath{\Xires^-}}\xspace}

\def\Omegares    {{\ensuremath{\POmega}}\xspace}

\def\Omegam      {{\ensuremath{\Omegares^-}}\xspace}

\def\Xicp        {{\ensuremath{\Xires^+_\cquark}}\xspace}

\def\Lb           {{\ensuremath{\Lz^0_\bquark}}\xspace}

\def\Xibm         {{\ensuremath{\Xires^-_\bquark}}\xspace}

\newcommand{\decay}[2]{\ensuremath{#1\!\to #2}\xspace} 

\def\to                 {\ensuremath{\rightarrow}\xspace}

\def\CP                {{\ensuremath{C\!P}}\xspace}

\def\AT#1     {\ensuremath{A_{\mathrm{T}}^{#1}}\xspace}           

\def\C#1      {\ensuremath{\mathcal{C}_{#1}}\xspace}                       
\def\Cp#1     {\ensuremath{\mathcal{C}_{#1}^{'}}\xspace}                    
\def\Ceff#1   {\ensuremath{\mathcal{C}_{#1}^{\mathrm{(eff)}}}\xspace}        
\def\Cpeff#1  {\ensuremath{\mathcal{C}_{#1}^{'\mathrm{(eff)}}}\xspace}       
\def\Ope#1    {\ensuremath{\mathcal{O}_{#1}}\xspace}                       
\def\Opep#1   {\ensuremath{\mathcal{O}_{#1}^{'}}\xspace}                    

\newcommand{\nospaceunit}[1]{\ensuremath{\text{#1}}}       
\newcommand{\aunit}[1]{\ensuremath{\text{\,#1}}}       

\newcommand{\tev}{\aunit{Te\kern -0.1em V}\xspace}
\newcommand{\gev}{\aunit{Ge\kern -0.1em V}\xspace}
\newcommand{\mev}{\aunit{Me\kern -0.1em V}\xspace}
\newcommand{\kev}{\aunit{ke\kern -0.1em V}\xspace}
\newcommand{\ev}{\aunit{e\kern -0.1em V}\xspace}
 
\newcommand{\mevc}{\ensuremath{\aunit{Me\kern -0.1em V\!/}c}\xspace}
\newcommand{\gevc}{\ensuremath{\aunit{Ge\kern -0.1em V\!/}c}\xspace}
\newcommand{\mevcc}{\ensuremath{\aunit{Me\kern -0.1em V\!/}c^2}\xspace}
\newcommand{\gevcc}{\ensuremath{\aunit{Ge\kern -0.1em V\!/}c^2}\xspace}

\def\cm   {\aunit{cm}\xspace}

\def\mm   {\aunit{mm}\xspace}

\def\mum  {\ensuremath{\,\upmu\nospaceunit{m}}\xspace}

\def\fb   {\ensuremath{\aunit{fb}}\xspace}
\def\invfb   {\ensuremath{\fb^{-1}}\xspace}

\def\ns   {\ensuremath{\aunit{ns}}\xspace}
\def\ps   {\ensuremath{\aunit{ps}}\xspace}

\def\gsim{{~\raise.15em\hbox{$>$}\kern-.85em
          \lower.35em\hbox{$\sim$}~}\xspace}
\def\lsim{{~\raise.15em\hbox{$<$}\kern-.85em
          \lower.35em\hbox{$\sim$}~}\xspace}

\def\sqs   {\ensuremath{\protect\sqrt{s}}\xspace}

\def\pt         {\ensuremath{p_{\mathrm{T}}}\xspace}

\def\ptot       {\ensuremath{p}\xspace}

\def\mrad{\aunit{mrad}\xspace}

\def\evtgen     {\mbox{\textsc{EvtGen}}\xspace}

\def\geant      {\mbox{\textsc{Geant4}}\xspace}

\def\photos     {\mbox{\textsc{Photos}}\xspace}

\def\pythia     {\mbox{\textsc{Pythia}}\xspace}

\def\tell1  {TELL1\xspace}
\def\ukl1   {UKL1\xspace}

\newcommand{\ie}{\mbox{\itshape i.e.}\xspace}

\newcommand{\lhcborcid}[1]{\href{https://orcid.org/#1}{\hspace*{0.1em}\raisebox{-0.45ex}{\includegraphics[width=1em]{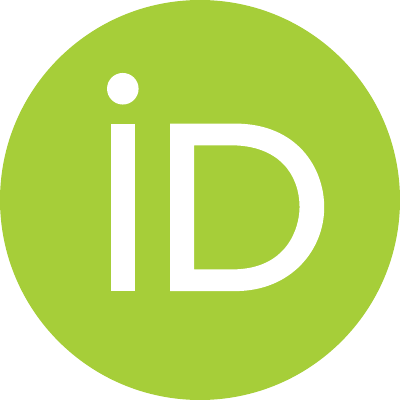}}}}

\def\npv         {{\ensuremath{n_{\text{PV}}}\xspace}}

\usepackage{cite} 
\usepackage{mciteplus}

\usepackage{longtable,adjustbox,amsmath} 
\usetikzlibrary{patterns,plotmarks,spy,arrows.meta,shapes.geometric}
\pgfdeclarelayer{fg}    
\pgfsetlayers{main,fg}
\begin{document}

\renewcommand{\thefootnote}{\fnsymbol{footnote}}
\setcounter{footnote}{1}


\begin{titlepage}
\pagenumbering{roman}

\vspace*{-1.5cm}
\centerline{\large EUROPEAN ORGANIZATION FOR NUCLEAR RESEARCH (CERN)}
\vspace*{1.5cm}
\noindent
\begin{tabular*}{\linewidth}{lc@{\extracolsep{\fill}}r@{\extracolsep{0pt}}}
\ifthenelse{\boolean{pdflatex}}
{\vspace*{-1.5cm}\mbox{\!\!\!\includegraphics[width=.14\textwidth]{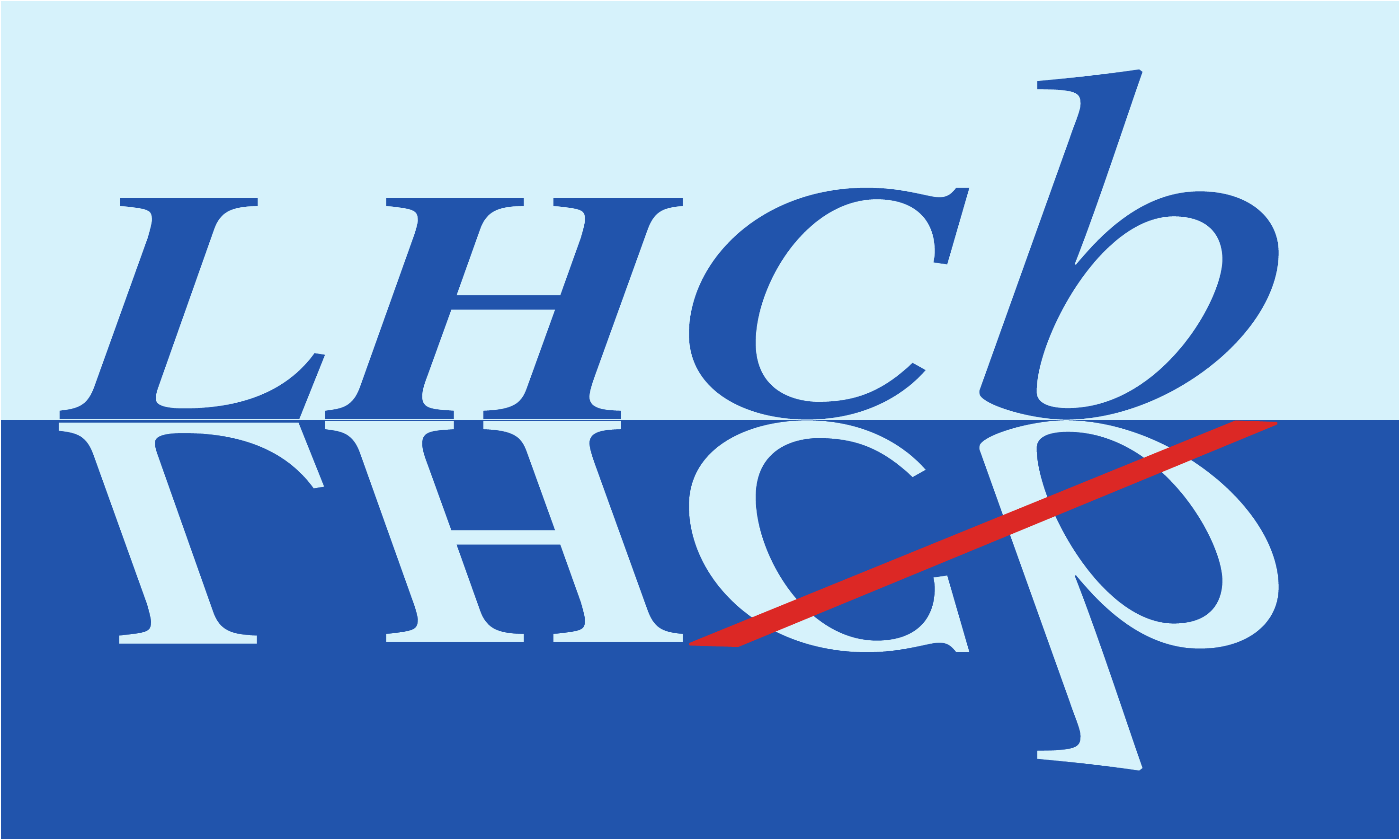}} & &}%
{\vspace*{-1.2cm}\mbox{\!\!\!\includegraphics[width=.12\textwidth]{figs/lhcb-logo.eps}} & &}%
\\
 & & CERN-EP-2024-077 \\  
 & & LHCb-DP-2023-004 \\  
 & & September 18, 2024 \\ 
 & & \\
\end{tabular*}

\vspace*{4.0cm}

{\normalfont\bfseries\boldmath\huge
\begin{center}
  \papertitle 
\end{center}
}

\vspace*{2.0cm}

\begin{center}
\paperauthors\footnote{Authors are listed at the end of this paper.}
\end{center}

\vspace{\fill}

\begin{abstract}
  \noindent
  A method is presented to reconstruct charged particles with lifetimes between $10\ps$ and $10\ns$, which considers a combination of their decay products and the partial tracks created by the initial charged particle. Using the $\Xim$ baryon as a benchmark, the method is demonstrated with simulated events and proton-proton collision data at $\sqs=13\tev$, corresponding to an integrated luminosity of 2.0\invfb collected with the \lhcb detector in 2018. Significant improvements in the angular resolution and the signal purity are obtained. The method is implemented as part of the LHCb Run~3 event trigger in a set of requirements to select detached hyperons. This is the first demonstration of the applicability of this approach at the LHC, and the first to show its scaling with instantaneous luminosity. 
  
\end{abstract}

\vspace*{2.0cm}

\begin{center}
Published in EPJC 84 (2024) 761
\end{center}

\vspace{\fill}

{\footnotesize 
\centerline{\copyright~\papercopyright. \href{\paperlicenceurl}{\paperlicence}.}}
\vspace*{2mm}

\end{titlepage}


\newpage
\setcounter{page}{2}
\mbox{~}


\renewcommand{\thefootnote}{\arabic{footnote}}
\setcounter{footnote}{0}

\cleardoublepage


\pagestyle{plain} 
\setcounter{page}{1}
\pagenumbering{arabic}


\section{Introduction}
\label{sec:Introduction}

Today's large-scale experiments at the LHC are designed to efficiently reconstruct the trajectories of charged particles that traverse the entire tracking system, such as pions, kaons, and muons. Despite having a finite lifetime, these particles can be treated as practically stable due to the significant boost they experience upon production at the LHC. Short-lived particles, with lifetimes $\tau<10\ps$, are exclusively reconstructed through the identification of their decay products.
The purpose of this article is to demonstrate a method for reconstructing charged particles which are neither short- nor very long-lived: specifically particles with a lifetime in the range of $10\ps$ to $10\ns$ and sufficient momentum to travel more than $10\cm$ before their decay. 

The interest in such particles, subsequently referred to as long-lived charged particles, originates on the one hand from the search for hypothetical beyond the Standard Model (SM) particles, that are produced on shell~\cite{Fairbairn:2006gg,Alimena:2019zri}. 
On the other hand, the reconstruction of long-lived particles provides an opportunity to study SM processes involving (long-lived) charged hyperons, which are ground-state baryons containing one or more strange quarks and no heavy quarks.
Notably, the $\Sigmap$, $\Sigmam$, $\Omegam$, and $\Xim$ baryons\footnote{The inclusion of charge conjugate processes and the use of natural units are implicit throughout this article.} are charged and have sufficiently long lifetimes to travel over $50\cm$ from their production vertex under typical LHC production conditions. 
Research involving these particles includes, but is not limited to, the precise study of \CP violation and rare decays~\cite{AlvesJunior:2018ldo,Alonso-Alvarez:2021oaj}, which are sensitive to physics beyond the SM, as well as hadron spectroscopy.

The LHCb experiment at the LHC is designed to study $\bquark$ and $\cquark$ hadrons. 
It is equipped with a vertex detector (\velo) that operates close to the beam axis. 
While this configuration is optimal for the study of heavy flavour hadrons, long-lived particles can escape the VELO before decaying. 
Consequently, only a fraction of the decay products can be reconstructed with the information of the VELO detector. 
A fraction of the decays can be recovered using tracks reconstructed with hits of tracking detectors downstream of the VELO, albeit with a worsened resolution. 
The reconstruction of long-lived charged particles may benefit from the information of hits in the VELO produced prior to the decay. 
This approach has been very fruitful at experiments outside of the LHC. 
The (fixed-target) FOCUS and E687 experiments~\cite{FOCUS,E-687:1990ses} employed this method to study charged hyperons as part of their physics programmes. 
Furthermore, the HyperCP experiment~\cite{HyperCP:2004kbv} was tailored to the tracking of these long-lived particles. 
The proposed ALICE 3 experiment at the LHC includes design elements that capture signals of strange particles before they decay~\cite{ALICE:2022wwr}.

Within LHCb, the initial interest in this technique originated from a different perspective: there is a long-standing need for new, precise measurements of the $\Kp$ meson mass, as the world-best measurements are not in agreement~\cite{PDG2022,Contu:1693666}. 
Subsequent to this study, using the $\decay{\Kp}{\pim\pip\pip}$ decay mode, several adaptations of these reconstruction techniques have been proposed for broader physics studies involving hyperons~\cite{Brochu:2016uiu}. 

Out of the charged particles with nanosecond lifetimes, the $\Xim$ baryon forms an excellent laboratory to understand how the information in the VELO detector can be beneficial, for two main reasons.
Firstly, it is abundantly produced at the LHC in different production mechanisms. 
Secondly, and most importantly, it predominantly decays to particles which leave signatures in the downstream tracking detectors. 
This property allows a direct comparison to the conventional reconstruction algorithms available at \lhcb. 

This article presents the performance benefits of tracking charged hyperons in the \lhcb vertex detector before they decay, a procedure referred to as \emph{VELO matching}, applied to the $\Xim$ baryon. 
The outcomes are derived from data collected by the \lhcb experiment during Run 2 of the LHC. Subsequently, this reconstruction technique is implemented within the \lhcb trigger for Run~3.
This manuscript closes with a discussion on other use-cases and potential extensions of the method. 
It is the first to demonstrate the performance of this tracking approach in LHC collision data, and assess its scalability for experiments conducted at higher instantaneous luminosity.

\section{Detector and simulation}
\label{sec:detector}
\begin{figure}[!t]
    \centering
    \includegraphics[width=0.85\textwidth]{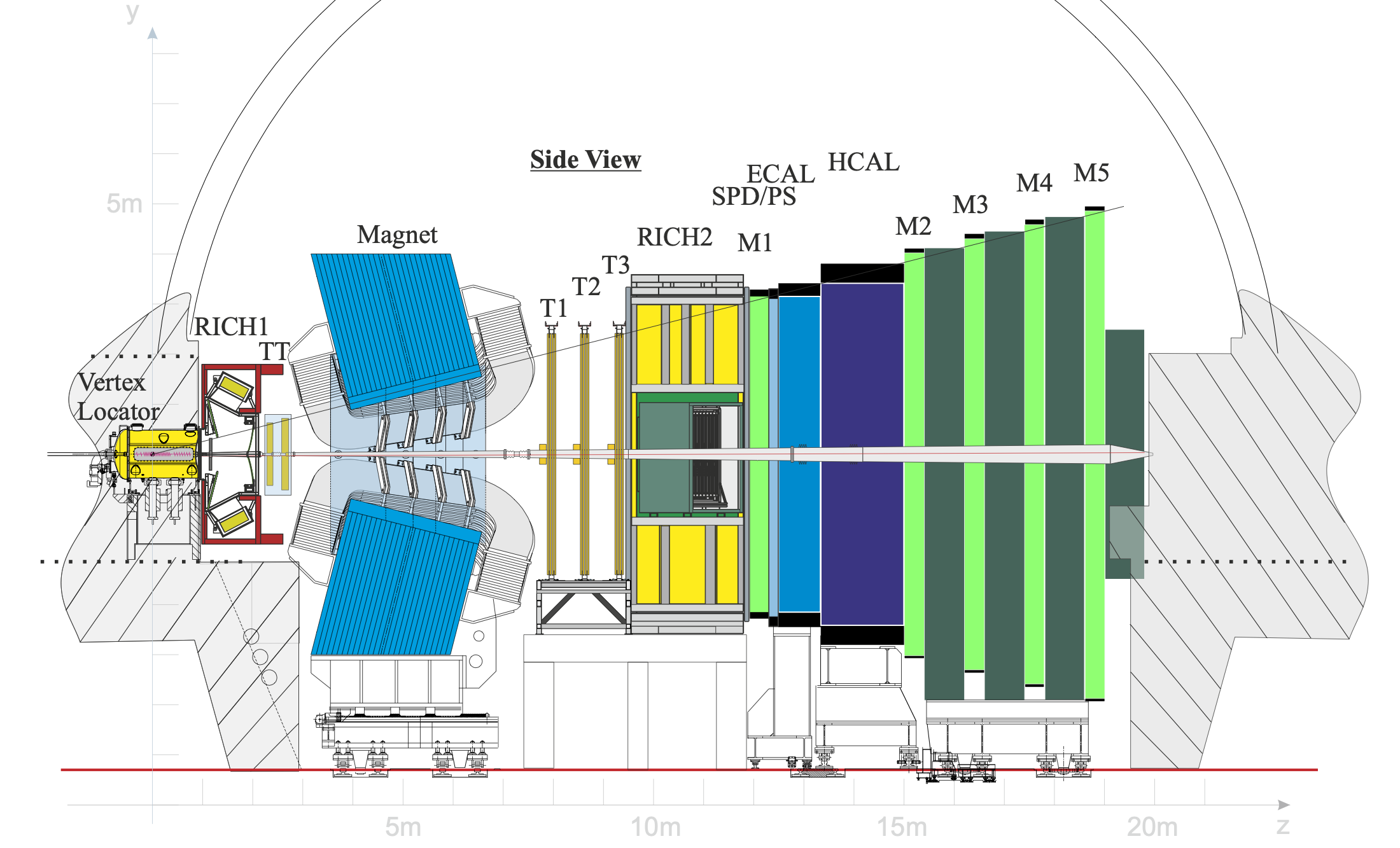}
    \caption{Schematic side-view of the LHCb detector as it operated during Runs 1 and 2, which shows the positioning of the various subdetectors. The horizontal axis in this picture corresponds to the beam axis. }
    \label{fig:lhcb_detector}
\end{figure}
The \lhcb detector~\cite{LHCb-DP-2008-001,LHCb-DP-2014-002} that operated during Runs 1 and 2 of the LHC is
 a single-arm forward spectrometer covering the \mbox{pseudorapidity} range $2<\eta <5$. A schematic of the detector is shown in Fig.~\ref{fig:lhcb_detector}. 
It includes a high-precision tracking system
consisting of a silicon-strip vertex detector, the aforementioned \velo, surrounding the $pp$
interaction region~\cite{LHCb-DP-2014-001}; a large-area silicon-strip detector, the TT, located
upstream of a dipole magnet with a bending power of about
$4{\mathrm{\,Tm}}$; and three stations of silicon-strip detectors (Inner Tracker) and straw
drift tubes (Outer Tracker)~\cite{LHCb-DP-2017-001} placed downstream of the magnet, referred to as the T stations.
The complete tracking system provides a measurement of the momentum, \ptot, of charged particles with
a relative uncertainty that varies from 0.5\% at low momentum to 1.0\% at 200\gev. Different types of charged hadrons are distinguished using information
from two ring-imaging Cherenkov (RICH) detectors~\cite{LHCb-DP-2012-003}. 
The impact parameter (IP) corresponds to the minimum distance of a track to a reconstructed $pp$ interaction vertex, referred to as a primary vertex (PV), and is measured with a resolution of $(15+29/\pt)\mum$,
where \pt is the component of the momentum transverse to the beam, in\,\gev. 

\begin{figure}[]\centering
	 \begin{tikzpicture}[scale=.11]
	   \tikzset{
  veloStation/.style={double, double distance=.1mm}
, puStation/.style={}
, ttLayer/.style={}
, magnet/.style={line width=2pt,draw=black!40!white,rounded corners=3mm}
, itLayer/.style={}
, otLayer/.style={}
, rich/.style={draw=none,fill=green!40!white}
, spd/.style={draw=none,fill=blue!40!white}
, converter/.style={draw=none,fill=green!40!white}
, prs/.style={draw=none,fill=yellow!40!white}
, ecal/.style={draw=none,fill=orange!40!white}
, hcal/.style={draw=none,fill=yellow!40!white}
, muon/.style={draw=none,fill=red!40!white}
}

\draw[veloStation] (-17.5mm,-0.2mm) -- (-17.5mm,5.0mm);
\draw[veloStation] (-14.5mm,-0.2mm) -- (-14.5mm,5.0mm);
\draw[veloStation] (-11.5mm,-0.2mm) -- (-11.5mm,5.0mm);
\draw[veloStation] (-8.5mm,-0.2mm) -- (-8.5mm,5.0mm);
\draw[veloStation] (-5.5mm,-0.2mm) -- (-5.5mm,5.0mm);
\draw[veloStation] (-2.5mm,-0.2mm) -- (-2.5mm,5.0mm);
\draw[veloStation] (0.5mm,-0.2mm) -- (0.5mm,5.0mm);
\draw[veloStation] (3.5mm,-0.2mm) -- (3.5mm,5.0mm);
\draw[veloStation] (6.5mm,-0.2mm) -- (6.5mm,5.0mm);
\draw[veloStation] (9.5mm,-0.2mm) -- (9.5mm,5.0mm);
\draw[veloStation] (12.5mm,-0.2mm) -- (12.5mm,5.0mm);
\draw[veloStation] (15.5mm,-0.2mm) -- (15.5mm,5.0mm);
\draw[veloStation] (18.5mm,-0.2mm) -- (18.5mm,5.0mm);
\draw[veloStation] (21.5mm,-0.2mm) -- (21.5mm,5.0mm);
\draw[veloStation] (24.5mm,-0.2mm) -- (24.5mm,5.0mm);
\draw[veloStation] (27.5mm,-0.2mm) -- (27.5mm,5.0mm);
\draw[veloStation] (43.5mm,-0.2mm) -- (43.5mm,5.0mm);
\draw[veloStation] (58.5mm,-0.2mm) -- (58.5mm,5.0mm);
\draw[veloStation] (63.5mm,-0.2mm) -- (63.5mm,5.0mm);
\draw[veloStation] (68.5mm,-0.2mm) -- (68.5mm,5.0mm);
\draw[veloStation] (73.5mm,-0.2mm) -- (73.5mm,5.0mm);
\draw[veloStation] (-16.0mm,0.2mm) -- (-16.0mm,-5.0mm);
\draw[veloStation] (-13.0mm,0.2mm) -- (-13.0mm,-5.0mm);
\draw[veloStation] (-10.0mm,0.2mm) -- (-10.0mm,-5.0mm);
\draw[veloStation] (-7.0mm,0.2mm) -- (-7.0mm,-5.0mm);
\draw[veloStation] (-4.0mm,0.2mm) -- (-4.0mm,-5.0mm);
\draw[veloStation] (-1.0mm,0.2mm) -- (-1.0mm,-5.0mm);
\draw[veloStation] (2.0mm,0.2mm) -- (2.0mm,-5.0mm);
\draw[veloStation] (5.0mm,0.2mm) -- (5.0mm,-5.0mm);
\draw[veloStation] (8.0mm,0.2mm) -- (8.0mm,-5.0mm);
\draw[veloStation] (11.0mm,0.2mm) -- (11.0mm,-5.0mm);
\draw[veloStation] (14.0mm,0.2mm) -- (14.0mm,-5.0mm);
\draw[veloStation] (17.0mm,0.2mm) -- (17.0mm,-5.0mm);
\draw[veloStation] (20.0mm,0.2mm) -- (20.0mm,-5.0mm);
\draw[veloStation] (23.0mm,0.2mm) -- (23.0mm,-5.0mm);
\draw[veloStation] (26.0mm,0.2mm) -- (26.0mm,-5.0mm);
\draw[veloStation] (29.0mm,0.2mm) -- (29.0mm,-5.0mm);
\draw[veloStation] (45.0mm,0.2mm) -- (45.0mm,-5.0mm);
\draw[veloStation] (60.0mm,0.2mm) -- (60.0mm,-5.0mm);
\draw[veloStation] (65.0mm,0.2mm) -- (65.0mm,-5.0mm);
\draw[veloStation] (70.0mm,0.2mm) -- (70.0mm,-5.0mm);
\draw[veloStation] (75.0mm,0.2mm) -- (75.0mm,-5.0mm);
\draw[puStation] (-31.5mm,-0.2mm) -- (-31.5mm,5.0mm);
\draw[puStation] (-23.5mm,-0.2mm) -- (-23.5mm,5.0mm);
\draw[puStation] (-30.0mm,0.2mm) -- (-30.0mm,-5.0mm);
\draw[puStation] (-22.0mm,0.2mm) -- (-22.0mm,-5.0mm);


\draw[ttLayer] (232.75mm,-85.6mm) -- (232.75mm,85.6mm);
\draw[ttLayer] (237.25mm,-85.6mm) -- (237.25mm,85.6mm);
\draw[ttLayer] (259.75mm,-85.6mm) -- (259.75mm,85.6mm);
\draw[ttLayer] (264.25mm,-85.6mm) -- (264.25mm,85.6mm);

\draw[magnet] (300.mm,200.mm) -- (750.mm,340.mm) -- (750.mm,-340.mm) -- (300.mm,-200.mm) -- cycle;

\draw[itLayer] (775.5mm,-62.1mm) -- (775.5mm,62.1mm); 
\draw[itLayer] (776.9mm,-64.1mm) -- (776.9mm,64.1mm); 
\draw[itLayer] (778.8mm,-64.1mm) -- (778.8mm,64.1mm); 
\draw[itLayer] (780.3mm,-62.1mm) -- (780.3mm,62.1mm); 
\draw[itLayer] (843.7mm,-62.8mm) -- (843.7mm,62.8mm); 
\draw[itLayer] (845.1mm,-64.8mm) -- (845.1mm,64.8mm); 
\draw[itLayer] (847.0mm,-64.8mm) -- (847.0mm,64.8mm); 
\draw[itLayer] (848.5mm,-62.8mm) -- (848.5mm,62.8mm); 
\draw[itLayer] (912.2mm,-63.5mm) -- (912.2mm,63.5mm); 
\draw[itLayer] (913.6mm,-65.5mm) -- (913.6mm,65.5mm); 
\draw[itLayer] (915.5mm,-65.5mm) -- (915.5mm,65.5mm); 
\draw[itLayer] (917.0mm,-63.5mm) -- (917.0mm,63.5mm); 

\draw[otLayer] (786.080130227mm,-322.0mm) -- (786.080130227mm,322.0mm);
\draw[otLayer] (791.530094892mm,-322.0mm) -- (791.530094892mm,322.0mm);
\draw[otLayer] (798.080196464mm,-322.0mm) -- (798.080196464mm,322.0mm);
\draw[otLayer] (803.530161128mm,-322.0mm) -- (803.530161128mm,322.0mm);
\draw[otLayer] (854.280588294mm,-322.0mm) -- (854.280588294mm,322.0mm);
\draw[otLayer] (859.730552959mm,-322.0mm) -- (859.730552959mm,322.0mm);
\draw[otLayer] (866.280654531mm,-322.0mm) -- (866.280654531mm,322.0mm);
\draw[otLayer] (871.730619195mm,-322.0mm) -- (871.730619195mm,322.0mm);
\draw[otLayer] (922.781008406mm,-322.0mm) -- (922.781008406mm,322.0mm);
\draw[otLayer] (928.230973071mm,-322.0mm) -- (928.230973071mm,322.0mm);
\draw[otLayer] (934.781110653mm,-322.0mm) -- (934.781110653mm,322.0mm);
\draw[otLayer] (940.231075317mm,-322.0mm) -- (940.231075317mm,322.0mm);


\tikzset{track/.style={line width=.5pt}}
\draw[track] (0mm,0mm) -- node[below left] {VELO track} (150mm,-25mm);
\draw[track] (200mm,55mm) -- (300mm,65mm) .. controls (500mm,80mm) .. (750mm,125mm) node[above left] {Downstream track} -- (1000mm,170mm);
\draw[track] (0mm,0mm) -- (300mm,-7.5mm) .. controls (500mm,-12.5mm) .. node[below] {Long track} (750mm,-30mm) -- (1000mm,-47.5mm);

\node[red] at (0,-370mm) {VELO};
\node[red] at (250mm,-370mm) {TT};
\node[red] at (520mm,-370mm) {Dipole magnet};
\node[red] at (866mm,-370mm) {T};
	 \end{tikzpicture}
	\caption{Schematic view of the different track types relevant for this paper, along with the used tracking detectors for these track types, showing the VELO detector, the TT detector, the magnet and the downstream tracking stations (T). The names of the tracking detectors along with the positioning of the magnet are indicated below the figure. }
	\label{fig:experiment_track_types}
\end{figure}
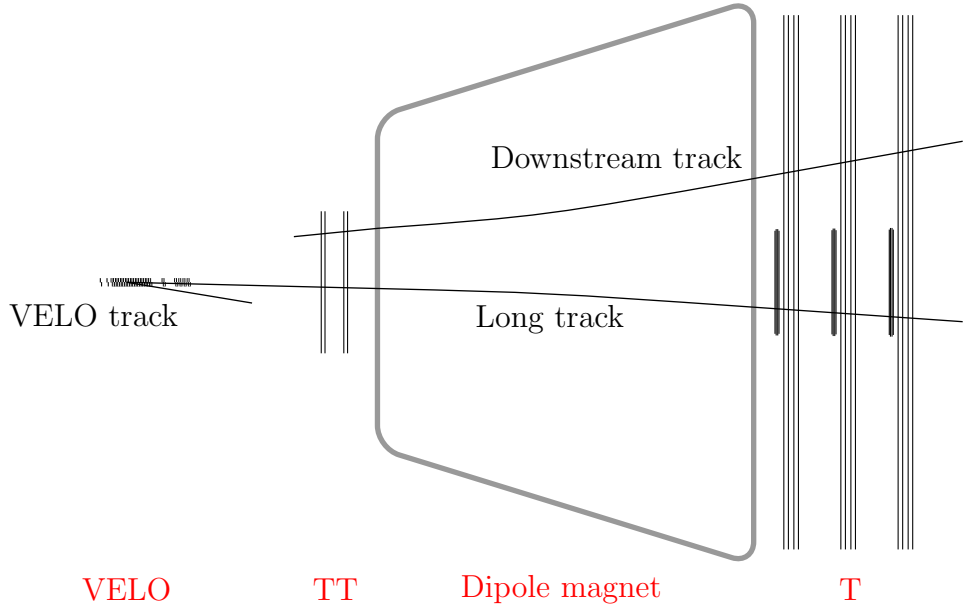

Figure~\ref{fig:experiment_track_types} shows an overview of the relevant track types used in this paper: \velo tracks, composed of at least three hits in the \velo; downstream tracks, which have hits in the T stations and the TT detector; and long tracks, which have hits in the VELO detector, T stations and optionally the TT detector. The momentum is inferred from the track's curvature in the magnetic field. Most physics analyses make use of long tracks, as they provide the best momentum and angular resolution among all of the track types. The \velo resides outside the magnetic field, and charged particles follow an approximately straight trajectory through the detector. Therefore, a \velo track only provides angular information about the particle's initial trajectory, and no information about the particle's absolute momentum. 

Simulated events are used to validate the proposed method. 
  In the simulation, $\proton\proton$ collisions are generated using
\pythia~\cite{Sjostrand:2007gs} 
 with a specific \lhcb
configuration~\cite{LHCb-PROC-2010-056}.  Decays of unstable particles
are described by \evtgen~\cite{Lange:2001uf}, in which final-state
radiation is generated using \photos~\cite{Golonka:2005pn}. The
interaction of the generated particles with the detector, and its response,
are implemented using the \geant
toolkit~\cite{Allison:2006ve, *Agostinelli:2002hh} as described in
Ref.~\cite{LHCb-PROC-2011-006}. 

\section{Candidate selection and VELO matching method}
\label{sec:selection}
\begin{figure}[!t]
  \begin{adjustbox}{width=\textwidth}    
    \input{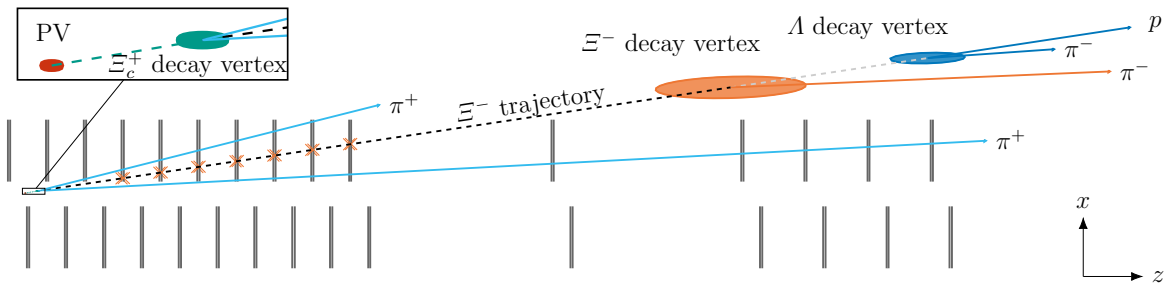}
  \end{adjustbox}  
  \caption[Velo sketch]{Illustration of the $\decay{\Xicp}{\Xim\pip\pip}$ decay. The VELO modules are drawn as vertical gray lines; reconstructed final state particles as colored straight lines; trajectories of decaying particles as dashed lines; decay vertices with their corresponding spatial uncertainty as colored ellipses; and the orange crosses represent hits from the $\Xim$ traversing the sensors.} \label{fig:velo_sketch}
\end{figure}

The $\Xim$ baryons are reconstructed as part of an abundant weak decay of the $\Xicp$ baryon, \mbox{$\decay{\Xicp}{\Xim\pip\pip}$}, which is used to benchmark the VELO matching method. The $\Xim$ baryon primarily decays to $\Lz \pim$, with \mbox{$\decay{\Lz}{\proton \pim}$}. These decays are selected in the $\sqrt{s}=13\tev$ $\proton\proton$ collision data recorded by the \lhcb detector in 2018, corresponding to an integrated luminosity of $2.0\invfb$. The reconstruction and selection procedure is split in two parts: a conventional selection, which uses downstream or long tracks to reconstruct the hyperons in the \lhcb software trigger, and an offline step which matches the result of the downstream reconstruction with a candidate $\Xim$-baryon track in the vertex detector, as illustrated in Fig.~\ref{fig:velo_sketch}. 

\subsection{Reconstruction and selection}
The online event selection is performed by a trigger~\cite{LHCb-DP-2019-001,LHCb-DP-2019-002}, 
which consists of a hardware stage, based on information from the calorimeter and muon systems, 
followed by two software stages, applying partial (HLT1) and full event reconstruction (HLT2).
While generic selections are used in the hardware and first software stage, dedicated selection algorithms for \mbox{$\decay{\Xicp}{\Xim\pip\pip}$} candidates, using both long- and downstream-track reconstruction, have been deployed for 2018 data taking. 
This dedicated path of selecting and reconstructing candidates contributes more than 70\% to the considered \mbox{$\decay{\Xicp}{\Xim\pip\pip}$} signal decays, in which the $\Xim$ is reconstructed from downstream tracks only.
The remaining 30\% is selected by trigger selections based on the topology of $b$-hadron decays~\cite{BBDT,LHCb-PROC-2015-018},
and an inclusive \mbox{$\decay{\Xim}{\Lz\pim}$} trigger selection, using only downstream tracks. 
Events passing the inclusive \mbox{$\decay{\Xim}{\Lz\pim}$} trigger selection are randomly discarded, such that on average one in 20 events is selected, in order to fit into the budgeted bandwidth of the LHCb triggers.

The dedicated $\Xim$ hyperon trigger selections start by building a $\Lz$ baryon candidate from two oppositely charged downstream tracks, which are of good quality\footnote{The quality of the tracks is evaluated through the differences between the reconstructed trajectory with the track's cluster positions. Similarly, for vertices the smallest distance between the reconstructed vertex and the contributing particles are used.} ($\chi^{2}_{\mathrm{track}}/\mathrm{ndf} < 4$) and combine to a vertex, applying modest quality requirements ($\chi^{2}_{\mathrm{vertex}}/\mathrm{ndf}<30$). These two tracks need to pass proton and pion particle-identification criteria, using 
combined information of the LHCb PID detectors~\cite{LHCb-DP-2018-001}. 
Another downstream track, assumed to be a pion, is combined with the $\Lz$-baryon candidate to 
reconstruct a $\Xim$ candidate. This combination is required to point towards the interaction region.
The dedicated \mbox{$\decay{\Xicp}{\Xim\pip\pip}$} trigger selection combines two displaced pions, reconstructed 
as long tracks with a significant IP with respect to any PV, with the $\Xim$ candidate to form a $\Xicp$ candidate. 
The candidate's vertex is required to have a good quality ($\chi^{2}_{\mathrm{vertex}}/\mathrm{ndf}<10$), be upstream of the $\Xim$ decay vertex, and point back to a PV.

\subsection{VELO matching algorithm}
\label{sec:selection_matching_algorithm}

After the reconstruction of $\Xim$ candidates from downstream tracks, 
a newly developed matching algorithm finds VELO tracks that are compatible with the $\Xim$ trajectory.
To simplify the computation initially, both the $\Xim$ trajectory and the VELO track are assumed to be straight 
lines, extrapolated from the $\Xim$ decay vertex and the last hit in the VELO.
The IP between the VELO track and the $\Xim$ decay vertex is required to be less than $5\mm$. In the majority of cases this requirement selects only one VELO track per $\Xim$ candidate.
Those VELO tracks are refitted with a Kalman filter using the $\Xim$ momentum, as reconstructed from downstream tracks, as input 
to obtain a more precise estimate of the VELO track parameters and their covariance matrix. 
In the track fit, the $\Xim$ candidate is extrapolated through the detector material 
and the fringe field of the dipole magnet to the position of the last hit recorded on the VELO track. 

As a last step, the algorithm re-computes the IP and a significance, $\chi^2_\mathrm{Match}$, for the 
matching procedure from the refitted VELO track and the transported $\Xim$ candidate:
\begin{equation*}
  \chi^2_\mathrm{Match} = 
    \left(\boldsymbol{s}_\text{Downstream} - \boldsymbol{s}_\mathrm{VELO}\right) 
    \left(\boldsymbol{\Sigma}_\text{Downstream} + \boldsymbol{\Sigma}_\mathrm{VELO}\right)^{-1} 
    \left(\boldsymbol{s}_\text{Downstream} - \boldsymbol{s}_\mathrm{VELO}\right)^\top,
\end{equation*}
with a vector of track parameters $\boldsymbol{s} = (x, y, t_x, t_y)$ and its covariance matrix $\boldsymbol{\Sigma}$; where $x,y$ are the positions and $t_x,t_y$ the slopes with respect to the beam axis. This vector is defined for both the VELO track and the extrapolated $\Xim$ candidate, reconstructed from downstream tracks, at the last VELO measurement.
It is possible to further refine the selection based on $\chi^2_\mathrm{Match}$, if needed. In the presented analysis, only a loose requirement is placed on the maximum matching $\chi^2_\mathrm{Match}$, which covers nearly all correct matches in simulated data. The $\chi^2_\mathrm{Match}$ also provides a ranking of possible multiple matches: in case there is more than one VELO track passing the IP selection, the one with the best $\chi^2_\mathrm{Match}$ is chosen.

\section{Performance}
\label{sec:data}
\begin{figure}[t]
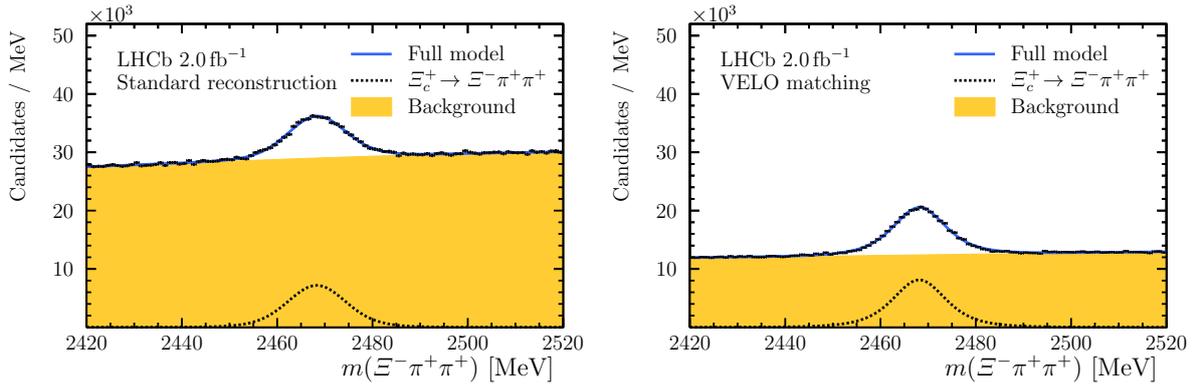

  \begin{adjustbox}{width=0.49\textwidth}
    \input{tikz_pictures/xipipi_DDD_18_selA_charm_binned}
  \end{adjustbox}
  \begin{adjustbox}{width=0.49\textwidth}    \input{tikz_pictures/xipipi_Match_18_selA_charm_binned}
  \end{adjustbox}
  \caption[Fits loose selection]{Invariant-mass distributions of $\decay{\Xicp}{\Xim\pip\pip}$ candidates, selected using typical online selection criteria, for (left) the standard reconstruction and (right) the VELO matching procedure. The fits, used to determine signal and background yields, are overlaid. }\label{fig:mass_online}
\end{figure}

The large resolution on the decay vertices of the $\Xim$ candidates in the conventional, or downstream, reconstruction makes their selection particularly susceptible to backgrounds.
One expects that only a small fraction of random combinations can be matched to a VELO track, such that applying the VELO matching procedure results in a reduction of these background contributions. 
Using the $\Xicp$ decays as benchmark, this is confirmed in data: Fig.~\ref{fig:mass_online} shows the distribution of the \mbox{$\Xim\pip\pip$} invariant mass, \mbox{$m(\Xim\pip\pip)$}, of candidates passing a loose, yet typical, trigger and offline selection for $\Xicp$ baryon decays as described in Sec.~\ref{sec:selection}. The right panel in the same figure presents the same distribution after the VELO matching procedure. 
The signal purity, calculated within the interval containing the central 95.45\% ($2\,\sigma$) of the signal probability density function, is improved from $12\%$ to $26\%$. 
The fitted signal yields suggest a large efficiency of the matching method, compatible with unity.
In simulation the efficiency of the matching algorithm varies between $96\%$ and $97\%$ depending on the kinematics of the $\Xim$ baryon, or whether the simulated $\Xim$ is produced promptly or from heavy flavor decays. 

\begin{figure}[t]
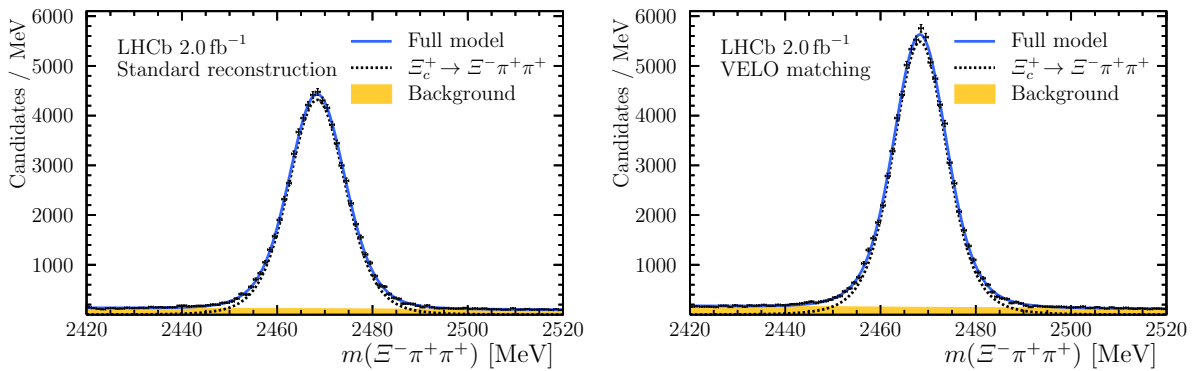

  \begin{adjustbox}{width=0.49\textwidth}
    \input{tikz_pictures/xipipi_DDD_18_charm_binned_bdt_gt_0p945}
  \end{adjustbox}
  \begin{adjustbox}{width=0.49\textwidth}    \input{tikz_pictures/xipipi_Match_18_charm_binned_bdt_gt_0p64}
  \end{adjustbox}
  \caption[Fits clean selection]{Invariant-mass distributions of $\decay{\Xicp}{\Xim\pip\pip}$ candidates with the full offline selection criteria for (left) the standard reconstruction, and (right) the VELO matching procedure. The fits, used to determine signal and background yields, are overlaid.}\label{fig:mass_fit_clean}
\end{figure}

A strict MVA-based~\cite{Hocker:2007ht} selection is subsequently applied to the same data set, using information related to particle identification, kinematics, and topological characteristics of the $\Xicp$ baryon. In the training of the classifier, simulated \mbox{$\decay{\Xicp}{\Xim\pip\pip}$} events are used as signal proxy, while candidates from the sideband region, \ie masses between 2220 and 2440\mev or 2500 and 2520\mev, are used as background proxy. The selection requirements are tuned separately for both reconstruction procedures, such that the same signal purity of $95\%$ is obtained in the invariant-mass distributions, as shown in Fig.~\ref{fig:mass_fit_clean}. The application of the matching procedure leads to a notable increase in the number of retained $\Xicp$ decays by $21\%$, demonstrating its benefit over the conventional reconstruction.

\subsection{Mass resolution}
\begin{figure}
    \centering
    \includegraphics[width=0.49\textwidth]{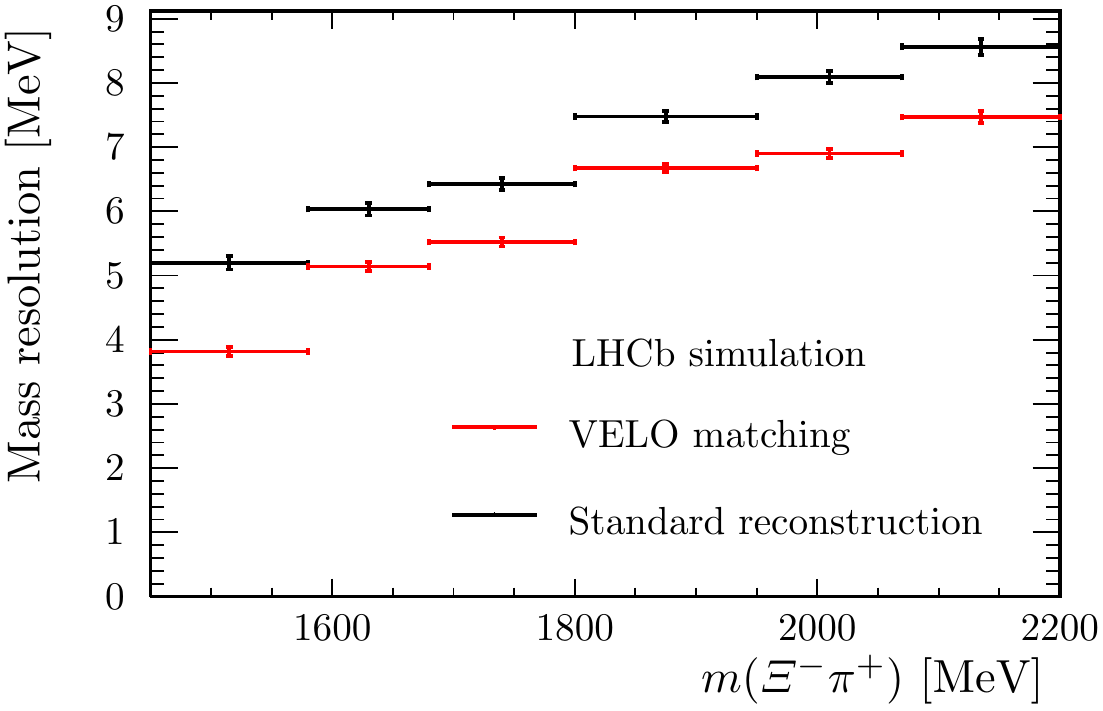}
    \includegraphics[width=0.49\textwidth]{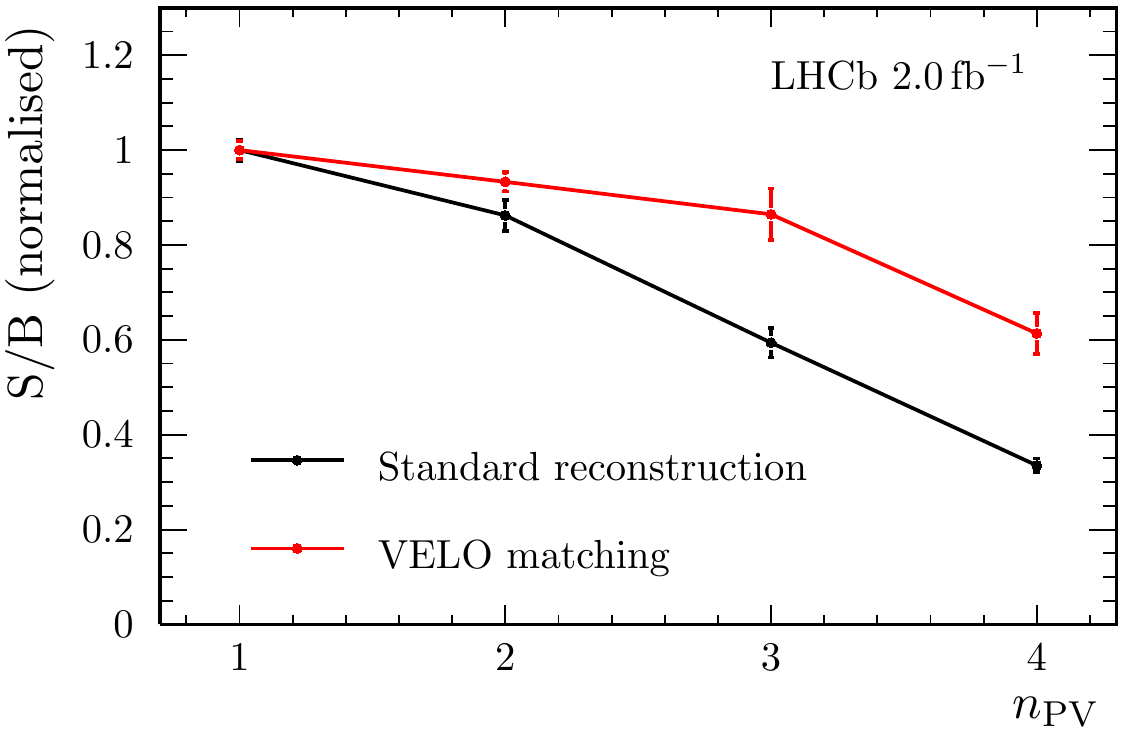}
    \caption{Left: Invariant-mass resolution of the $\Xim \pip$ combination in simulation for the standard reconstruction and the VELO matching. Right: Normalised signal-to-background ratio (S/B) in data for $\Xicp$ decays as a function of $n_{\text{PV}}$ for the standard, downstream-track reconstruction of the $\Xim$ baryon, and the reconstruction which additionally matches the $\Xim$ candidate to a VELO track. The points on each curve are normalised separately to their signal-to-background ratio at $n_{\text{PV}}=1$, to illustrate the differential behaviour.}
    \label{fig:mass_resolution_absolute}
\end{figure}

In addition to the resulting improved signal significance, an improved mass resolution is particularly valuable in the identification of resonances close to production thresholds and in accurately measuring  their properties.
A simulation study shows that the VELO-matching procedure improves the mass resolution of the $\Xim \pip$ pair in the $\decay{\Xicp}{\Xim\pip\pip}$ decay by over $1\mev$, or $25\%$, as presented in the left panel of Fig.~\ref{fig:mass_resolution_absolute}. The relative improvement is lower at higher masses of the $\Xim \pip$ system, as at higher masses the uncertainty on the momenta of the decay products is increasingly important. 

\subsection{Pile-up}
An upgrade of the LHCb experiment~\cite{LHCb-DP-2022-002}, the installation of which was completed in 2022, will take data for the rest of Run~3 of the LHC (2022--2025). 
During Run~3, LHCb operates at a higher instantaneous luminosity by increasing the average number of $\proton\proton$ interactions per beam crossing to about $5.2$, compared to about $1.1$ in 2018. 
If no new techniques are applied, the increase in the number of collisions is expected to lead to an increase of backgrounds in which the signal candidate is composed of tracks associated to a different primary interaction. 
This could be particularly problematic for trigger selections, in which criteria are kept to a minimum to avoid biasing the data.
Using the $\Xicp$ decay chain as benchmark, this potential decrease in the signal-to-background ratio is evaluated as a function of the number of PVs in the event, \npv. 
The data taken in 2018 include sufficient events with multiple PVs to evaluate the decrease in performance up to $\npv=4$. The applied selection criteria correspond to those in the HLT2 trigger selection, supplemented with a requirement on the \Xicp baryon flight direction being compatible with originating from a PV. 
The signal and background yields are determined through unbinned maximum-likelihood fits to the \mbox{$\Xim \pip \pip$} mass distributions. 
The full mass range, as implemented in the HLT2 trigger selection, is considered, as it determines the rate and the efficiency of the selection.
Since the signal purity is higher at $\npv=1$ for the VELO-matched $\Xim$ candidates, the differential behaviour is evaluated by first normalising the results to the signal-to-background ratio at $\npv=1$. 

The results are presented in the right panel of Fig.~\ref{fig:mass_resolution_absolute}. 
For the standard reconstruction, in $\npv=4$ events, the signal-to-background ratio deteriorates to only $(33.5 \pm 3.1)\%$ of its original value. 
However, when using the VELO matching, $(61.3 \pm 5.4)\%$ of its original value is retained. 
This result underlines the importance of implementing the VELO matching early in the selection for Run 3 of the LHC, ideally already in the software trigger, as it enables a significantly improved utilization of the limited available bandwidth.

\subsection{Identification of detached hyperon decays}\label{sec:ip_perf}
In the study of long-lived charged-particle decays, it is helpful to be able to distinguish the particles created in the primary interaction from those originating from heavy flavour decays. 
Because of the non-negligible lifetime of charm  and beauty hadrons, their decay products do not necessarily point to a primary interaction. 
For this reason, the IP is frequently used as the primary discriminating observable in the identification of particles produced in such decays. 

\begin{figure}
    \centering
    \includegraphics[width=0.50\textwidth]{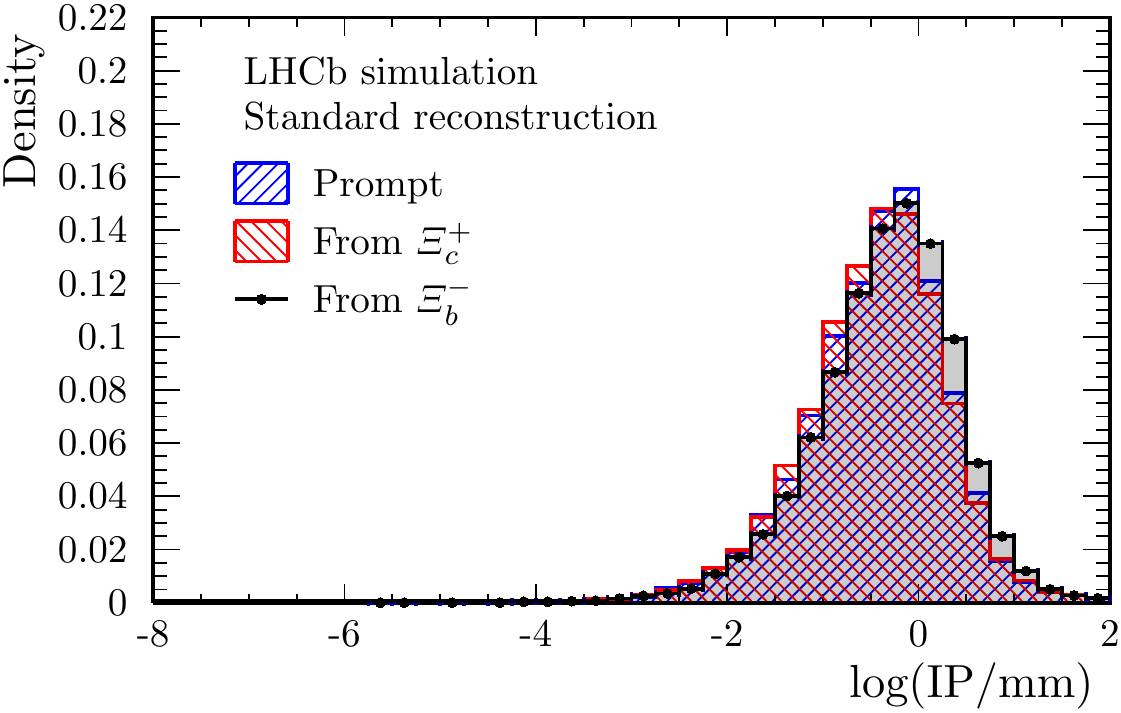}
    \includegraphics[width=0.48\textwidth]{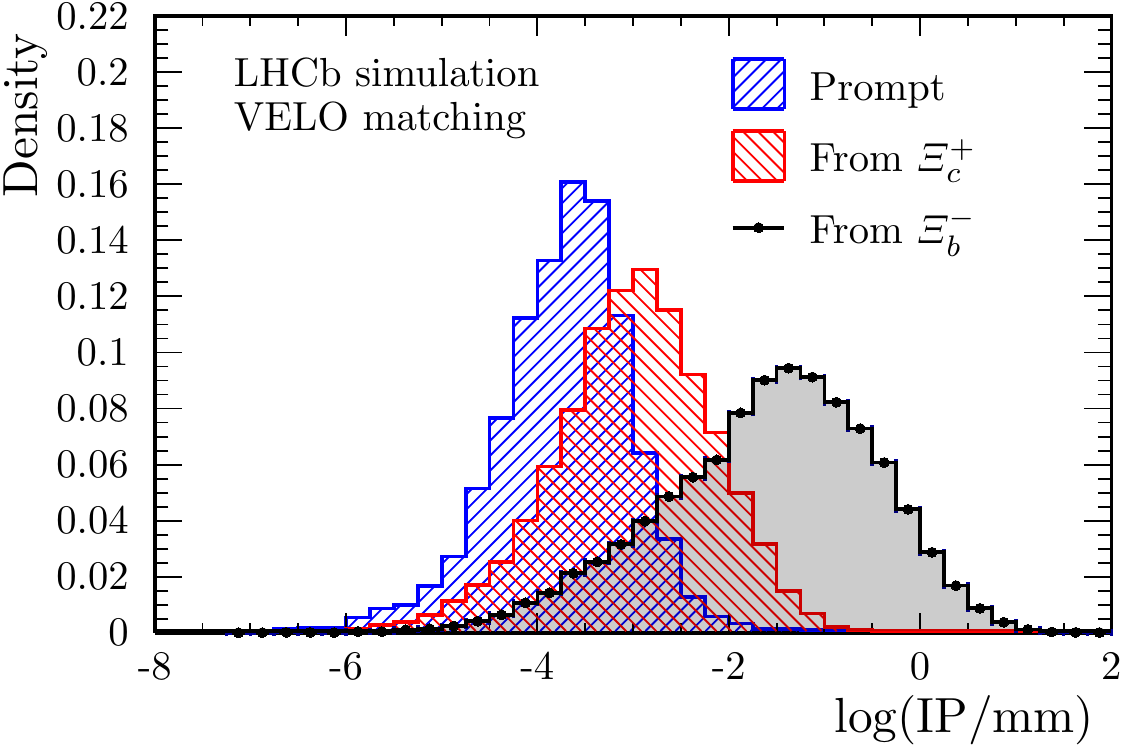} \\
    \includegraphics[width=0.50\textwidth]{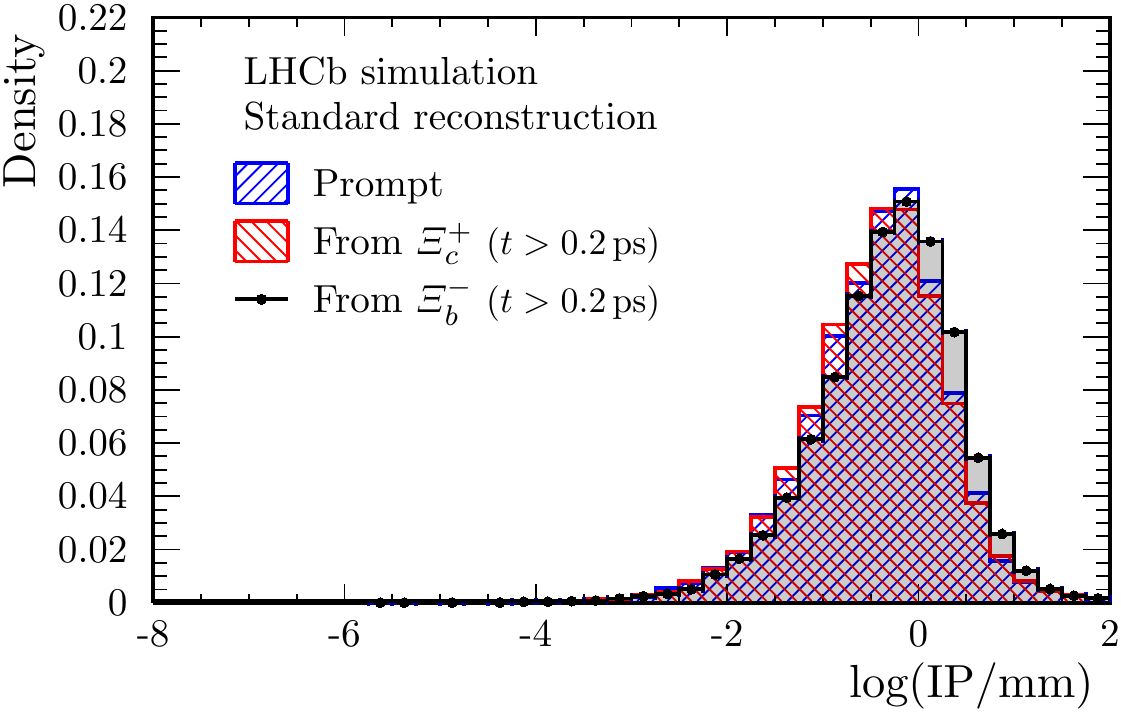}
    \includegraphics[width=0.48\textwidth]{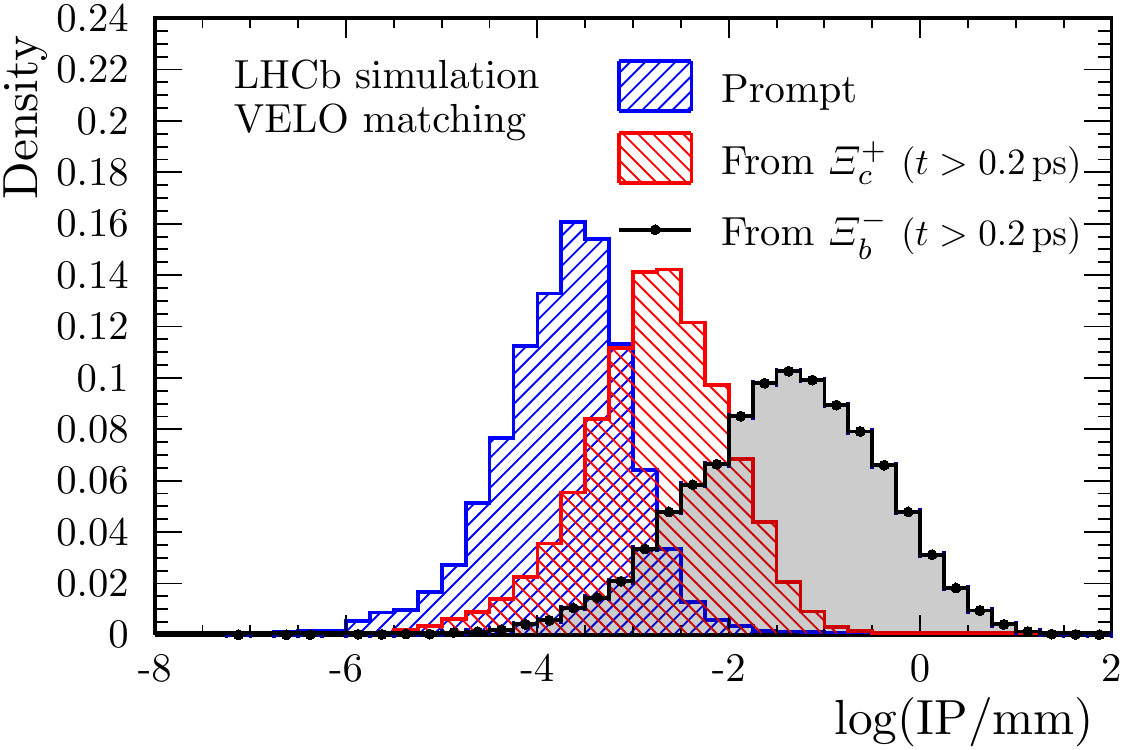}
    \caption{Distributions of the minimal natural logarithms of the IP, in mm, of the $\Xim$ baryon with respect to a PV, (left) with the standard reconstruction, (right) with the \velo matching method, (top) without a decay-time requirement, and (bottom) with a decay-time requirement $t>0.2\ps$. The ``from \Xicp'' (from \Xibm) component corresponds to simulated \mbox{$\decay{\Xicp}{\Xim\pip\pip}$} (\mbox{$\decay{\Xibm}{\jpsi\Xim}$}) decays. In each of the components, the $\Xim$ baryon is reconstructed through the \mbox{$\decay{\Xim}{\Lz\pim}$} decay. }
    \label{fig:ip_separation}
\end{figure}

Long-lived particles that are reconstructed using only downstream tracks have a considerably worse pointing resolution, corresponding to a resolution on the IP close to $1\mm$. 
This prevents the statistical separation between $\Xim$ baryons produced promptly, those via $\Xicp$ decays, and those from $\Xibm$ decays, as is shown in the top-left panel of Fig.~\ref{fig:ip_separation}.
Instead, as is seen in the top-right panel of this figure, the enhanced angular resolution of the $\Xim$ matched to a VELO track enables the separation of distinct production modes. The extended tails on the left side of the distribution pertaining to $\bquark$ and $\cquark$ hadron decays predominantly originate from the exponential nature of the decay-time distribution. This separation between the components is seen to increase already with a mild minimum decay-time requirement\footnote{This decay-time requirement has also been used in the optimisation of the LHCb event trigger~\cite{LHCb-PUB-2014-027}, and is representative for the events selected in a data analysis.}, $t > 0.2\ps$, as presented in the bottom panels of Fig.~\ref{fig:ip_separation}.

\section{Inclusive hyperon trigger selections}
\label{sec:run3_trigger}
During the course of Run 2 of the LHC, an inclusive trigger selection aimed at identifying $\Xim$ baryons from downstream tracks was implemented within the framework of LHCb's second-stage software trigger. This selection has been presented in Sec.~\ref{sec:selection}. However, to control the high rates of primarily prompt $\Xim$, including a considerable fraction of combinatorial background events, events passing this trigger selection are randomly discarded, such that on average only one in every 20 events is selected.

Following the presented performance improvements, the initial and computationally fast IP-based matching of the VELO track to the $\Xim$ candidate, discussed in Sec.~\ref{sec:selection_matching_algorithm}, has been implemented to run as part of the software trigger for the upgraded LHCb experiment.
In addition to the desired improvement in the signal-to-background ratio, the VELO matching provides a discrimination between prompt and secondary $\Xim$ baryons, as shown in Sec.~\ref{sec:ip_perf}. This enables the implementation of an inclusive event trigger for detached charged hyperons, focusing on $\Xim$ baryons produced in decays of $c$ and $b$ hadrons.
In this trigger selection, $\Xim$ candidates are required to have high transverse momentum and the matched VELO track, by a selection on the IP, is required to be displaced from any PV in the event.
For the Run 2 detector, these requirements remove about half of the prompt $\Xim$ baryons, while maintaining a high efficiency for decays from charm~($>93\%$) and beauty hadrons~($> 98\%$). For the Run 3 detector at a given pile-up, the performance is expected to be even better due to the improved resolution of the new, pixel-based vertex detector~\cite{LHCb-TDR-013}. This approach will select $\Xim$ baryons without having to discard events randomly. 
Following this strategy, a similar inclusive trigger selection has been implemented for detached $\Omegam$ baryons, reconstructed through its prominent $\decay{\Omegam}{\Lz\Km}$ decay.

\section{Outlook}
\label{sec:outlook}
The presented tracking procedure has many prospects for application within the LHCb physics programme. 
Concerning the Run 3 trigger, work is ongoing to implement the method in the partial event reconstruction in HLT1, where the reconstruction of downstream tracks has been added recently.
In the field of strange physics, the superior background rejection and angular information are also expected to be strong advantages throughout. The rare \mbox{$\decay{\Sigmap}{\proton \mup \mun}$} decay is a good example. The latest result of the LHCb collaboration~\cite{LHCb-PAPER-2017-049}, which is the world's best measurement, is obtained using only long tracks, as these provided a superior experimental resolution. With the VELO matching procedure, the considered dataset can be extended to include decays of the $\Sigmap$ baryon downstream of the vertex detector, more than doubling the expected data sample. 

In the introduction, the application of this technique to a measurement of the $\Kp$ meson mass was mentioned. The exploratory study~\cite{Contu:1693666} identified that the small opening angle inherent to the $\decay{\Kp}{\pim\pip\pip}$ decay requires additional attention, as signals from the $\Kp$ meson can be falsely associated to one of the decay products, and vice-versa. The presented technique can be further refined, taking such overlaps into consideration. This technique, together with the precise momentum resolution of the LHCb spectrometer, could allow for the desired, precise measurement of the $\Kp$ meson mass.

The precise angular resolution of the long-lived particle is expected to benefit analyses of decays with unreconstructed particles. An illustrative example is that of the semileptonic decay \mbox{$\decay{\Xim} {\Lz \mun \neumb}$}~\cite{Cabibbo:2003ea}, in which the neutrino is not reconstructed. Topological constraints, including the system's mass and direction information, enable the determination of the neutrino's momentum. This technique has been applied regularly in the context of heavy flavour decays~\cite{LHCB-PAPER-2018-013,LHCB-PAPER-2018-024}. The limiting resolution for heavy flavour decays hinges on the direction of the long-lived particle. If the direction of the hyperon is measured via a track, rather than inferred from the secondary and primary vertices as in the case of heavy flavour decays, a superior resolution is anticipated. Consequently, it is expected that decays with missing particles, such as these semileptonic decays, can be studied with improved precision. 

\begin{figure}[]
    \centering
  \begin{adjustbox}{width=0.67\textwidth}
    \input{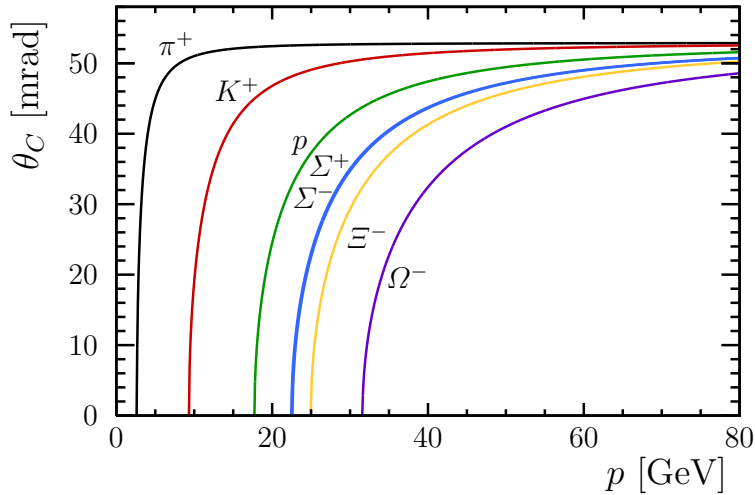}
  \end{adjustbox}
  \caption[Cherenkov angles RICH 1]{
  Cherenkov angles for different charged particle species traversing the RICH1, which uses $C_{4}F_{10}$ gas as radiator.}\label{fig:Cherenkov}
\end{figure}

The presented tracking procedure focuses on scenarios where the long-lived particle decays prior to reaching the TT detector, allowing it to be reconstructed as a VELO track, and its decay products as downstream tracks. 
In situations where the particle's flight distance extends further downstream, the decay products are not anticipated to be reconstructed as downstream tracks. Instead, the long-lived particle passes through both the VELO and the TT tracking detectors, and potentially the tracking detectors located downstream of the magnet. 
The particle traverses substantially more of the dipole magnet's field in this case, allowing for an accurate estimate of its momentum. This makes the reconstruction of the long-lived particle feasible without the need to reconstruct its decay products. 
To reject backgrounds of other, abundant, stable particles, it is possible to still search for signals of these decay products in the downstream tracking stations. 
In this scenario, another approach might prove promising: the particle-identification detectors could be used to help identify the particle species. Figure~\ref{fig:Cherenkov} presents the expected Cherenkov angles in the RICH detector upstream of the dipole magnet, RICH1, for the common particle species, along with several hyperons. 
For momenta below $80\gev$, the hyperons could be distinguished from the abundant pions, kaons and protons, when only accounting for the finite resolution of the RICH detectors of below $1~\mrad$~\cite{LHCb-TDR-014}. 
Studies with fast simulation~\cite{Cowan:2016tnm} show that this momentum range would cover over $95\%$ of the $\Xim$ baryons considered in the analysis of Sec.~\ref{sec:data}, and that more than 25\% of all $\Xim$ baryons in the LHCb acceptance will decay downstream of the TT detector. 
However, the momentum resolution of tracks reconstructed using only the VELO and TT information is limited. This  could potentially jeopardise the discrimination based on the RICH information between hyperons and more abundant particle species. 
Therefore, while recommending further exploration in this direction, it is anticipated that using additional information from the underlying decay topology will still be necessary to attain a satisfactory performance.

Conversely, the presented methodology can be extended to encompass decays that happen within the VELO detector. Taking the example of $\decay{\Xim}{\Lz\pim}$ decays mentioned earlier, both the $\Xim$ and $\pim$ could be reconstructed as distinct tracks with different slopes. This reconstruction would use information from the VELO detector to provide an adequate angular resolution. This scenario, or decay topology, is referred to as kink reconstruction. If these kinks can be resolved and isolated well, they form a starting point to reconstruct the decay of a charged particle into charged and neutral particles. Notable examples include the studied \mbox{$\decay{\Xim}{\Lz\pim}$} decay, but also charged pion and kaon decays, \mbox{$\decay{\Sigmap}{\proton\piz}$} decays, and yet-to-be-explored decays at LHCb such as the \mbox{$\decay{\Sigmam}{\neutron\pim}$} decay. Their reconstruction would enable the study of currently inaccessible charm and beauty decays, such as \mbox{$\decay{\Lb}{\Sigmam \jpsi \pip}$}. 
Related to kink topologies is the addition of single, isolated hits in the vicinity of a reconstructed decay vertex as constraints in the reconstruction of the decaying particle. 
Such a technique would not only work for decaying charged particles like the $\Xim$, but also $\KS$ meson and $\Lz$ baryon decays, whose decay products have traversed too few sensors to be reconstructed as VELO tracks.

\section{Conclusion and summary}
\label{sec:conclusion}
Using data recorded by the LHCb experiment in 2018, the reconstruction of partial tracks of $\Xim$ baryons in the vertex detector has demonstrated an improvement in capabilities to study long-lived charged particles at the LHC.

Using the \mbox{$\decay{\Xicp}{\Xim\pip\pip}$} decay as a benchmark, the performance of a newly implemented VELO matching algorithm was studied. The efficiency of the matching algorithm was found to vary between 96 and 97\% depending on the production mode and kinematics of the $\Xim$ baryon. Despite this small loss in efficiency, the VELO matching leads to an increase of more than 20\% in the number of retained $\decay{\Xicp}{\Xim\pip\pip}$ signal events compared to the standard \lhcb reconstruction, given a selection purity of 95\%. The mass resolution of the $\Xim\pip$ combination improves by 25\% and the signal-to-background ratio as a function of the number of PVs improves significantly. In view of Run 3 and beyond, the latter is particularly important, as it offers insights into how the efficacy of this reconstruction algorithm will extend to environments with higher instantaneous luminosities. Finally, it has been shown that the VELO matching method allows for clear discrimination of $\Xim$ baryons produced in heavy flavor decays as opposed to the standard reconstruction. This enabled the implementation of inclusive detached $\Xim$ and $\Omegam$ event triggers for Run 3 data taking at LHCb, which would have not been feasible with the standard reconstruction method given the trigger bandwidth constraints. These results  highlight the potential impact of tailored procedures to maximise the sensitivity of the LHC experiments in their searches for physics beyond the Standard Model.

\section*{Acknowledgements}
%
%
\noindent We express our gratitude to our colleagues in the CERN
accelerator departments for the excellent performance of the LHC. We
thank the technical and administrative staff at the LHCb
institutes.
We acknowledge support from CERN and from the national agencies:
CAPES, CNPq, FAPERJ and FINEP (Brazil); 
MOST and NSFC (China); 
CNRS/IN2P3 (France); 
BMBF, DFG and MPG (Germany); 
INFN (Italy); 
NWO (Netherlands); 
MNiSW and NCN (Poland); 
MCID/IFA (Romania); 
MICINN (Spain); 
SNSF and SER (Switzerland); 
NASU (Ukraine); 
STFC (United Kingdom); 
DOE NP and NSF (USA).
We acknowledge the computing resources that are provided by CERN, IN2P3
(France), KIT and DESY (Germany), INFN (Italy), SURF (Netherlands),
PIC (Spain), GridPP (United Kingdom), 
CSCS (Switzerland), IFIN-HH (Romania), CBPF (Brazil),
and Polish WLCG (Poland).
We are indebted to the communities behind the multiple open-source
software packages on which we depend.
Individual groups or members have received support from
ARC and ARDC (Australia);
Key Research Program of Frontier Sciences of CAS, CAS PIFI, CAS CCEPP, 
Fundamental Research Funds for the Central Universities, 
and Sci. \& Tech. Program of Guangzhou (China);
Minciencias (Colombia);
EPLANET, Marie Sk\l{}odowska-Curie Actions, ERC and NextGenerationEU (European Union);
A*MIDEX, ANR, IPhU and Labex P2IO, and R\'{e}gion Auvergne-Rh\^{o}ne-Alpes (France);
AvH Foundation (Germany);
ICSC (Italy); 
GVA, XuntaGal, GENCAT, Inditex, InTalent and Prog.~Atracci\'on Talento, CM (Spain);
SRC (Sweden);
the Leverhulme Trust, the Royal Society
 and UKRI (United Kingdom).

\newpage
\section*{Supplemental material}
\appendix

\section{Distribution of $\chi^2_\mathrm{Match}$}
\label{sec:dist_chi2_match}
\begin{figure}[b]
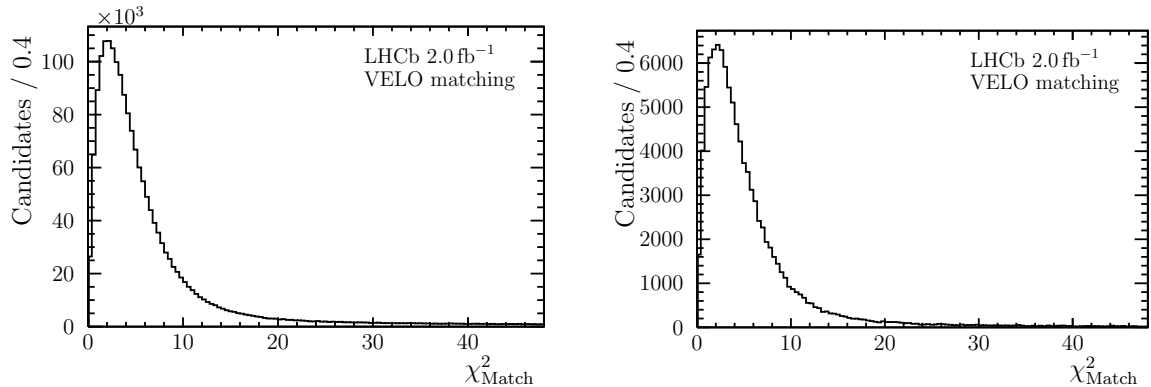

  \begin{adjustbox}{width=0.49\textwidth}
    \input{tikz_pictures/chi2_match_xipipi_Match_18_selA}
  \end{adjustbox}
  \begin{adjustbox}{width=0.49\textwidth}    \input{tikz_pictures/chi2_match_xipipi_Match_18_selA_bdt_gt_0p64}
  \end{adjustbox}
  \caption{Distributions of $\chi^2_\mathrm{Match}$ for data corresponding to (left) the online selection, as introduced in Fig.~\ref{fig:mass_online}, and (right) the offline selection, as introduced in Fig.~\ref{fig:mass_fit_clean}. No significant differences between the two distributions are expected, since the purity of $\Xim$ candidates entering the selection are similar, while the background in Fig.~\ref{fig:mass_online} mainly comes from combinations of a real $\Xim$ baryons with random pions.}\label{fig:chi2match}
\end{figure}

The distributions of the $\chi^2_\mathrm{Match}$ variable, introduced in Sec.~\ref{sec:selection_matching_algorithm}, are shown in Fig.~\ref{fig:chi2match}. The selection only considers matches up to $\chi^2_\mathrm{Match}=48$. 

\clearpage


\addcontentsline{toc}{section}{References}
\bibliographystyle{LHCb}
\bibliography{main,standard,LHCb-PAPER,LHCb-CONF,LHCb-DP,LHCb-TDR}

\newpage
\centerline
{\large\bf LHCb collaboration}
\begin
{flushleft}
\small
R.~Aaij$^{36}$\lhcborcid{0000-0003-0533-1952},
A.S.W.~Abdelmotteleb$^{55}$\lhcborcid{0000-0001-7905-0542},
C.~Abellan~Beteta$^{49}$,
F.~Abudin{\'e}n$^{55}$\lhcborcid{0000-0002-6737-3528},
T.~Ackernley$^{59}$\lhcborcid{0000-0002-5951-3498},
J. A. ~Adams$^{67}$\lhcborcid{0009-0003-9175-689X},
A. A. ~Adefisoye$^{67}$\lhcborcid{0000-0003-2448-1550},
B.~Adeva$^{45}$\lhcborcid{0000-0001-9756-3712},
M.~Adinolfi$^{53}$\lhcborcid{0000-0002-1326-1264},
P.~Adlarson$^{79}$\lhcborcid{0000-0001-6280-3851}
% ...

{\footnotesize \it

% ...
$^{36}$Nikhef National Institute for Subatomic Physics, Amsterdam, Netherlands\\
% ...
$^{45}$Instituto Galego de F{\'\i}sica de Altas Enerx{\'\i}as (IGFAE), Universidade de Santiago de Compostela, Santiago de Compostela, Spain\\
% ...
$^{49}$Physik-Institut, Universit{\"a}t Z{\"u}rich, Z{\"u}rich, Switzerland\\
% ...
$^{53}$H.H. Wills Physics Laboratory, University of Bristol, Bristol, United Kingdom\\
% ...
$^{55}$Department of Physics, University of Warwick, Coventry, United Kingdom\\
% ...
$^{58}$School of Physics and Astronomy, University of Glasgow, Glasgow, United Kingdom\\
$^{59}$Oliver Lodge Laboratory, University of Liverpool, Liverpool, United Kingdom\\
% ...
$^{67}$Syracuse University, Syracuse, NY, United States\\
% ...
$^{79}$Department of Physics and Astronomy, Uppsala University, Uppsala, Sweden, associated to $^{58}$\\
% ...
}
\end{flushleft}

\end{document}